\documentclass[a4paper]{article}

\usepackage{graphicx}
\usepackage{amsfonts}
\usepackage{amsmath}
\usepackage{amsbsy}
\usepackage{subfigure}
\usepackage{geometry}
\usepackage{setspace} 
\usepackage{bm}
\usepackage{multicol}
\usepackage{multirow}
\usepackage{color} 

\title{ Magnetic Braiding and Quasi-Separatrix Layers}
\author{A. L. Wilmot-Smith$^{\star}$, G. Hornig, D. I. Pontin. \\
{\small Division of Mathematics, University of Dundee, Dundee, DD1 4HN, UK}}

\begin{document}

\maketitle

\begin{abstract}
The squashing factor $Q$ (Titov {\it et al.} 2002), a property of the magnetic field line mapping,  
has been suggested as an indicator for the formation of current sheets, and subsequently
 magnetic reconnection, in astrophysical plasmas.  Here we test this hypothesis for a 
particular class of braided magnetic fields which serve as a model for solar coronal loops. 
We explore the relationship between Quasi-Separatrix Layers (QSLs), that is, layer-like 
structures with high $Q$ value, electric  currents and integrated parallel currents, the latter 
being a quantity closely related to the reconnection rate.  
It is found that as the degree of braiding of the magnetic field is increased  the maximum
values of $Q$ increase exponentially.   At the same time, the distribution of $Q$ becomes 
increasingly filamentary, with the width of the high-$Q$ layers exponentially 
decreasing. This is accompanied by an increase in the number of layers so
that as the field is increasingly braided the volume becomes occupied by a myriad 
of thin QSLs.   
QSLs are not found to be good predictors of current features in this class of braided fields.
Indeed, despite the presence of multiple QSLs, the current associated with the field remains 
smooth and large-scale under ideal relaxation; the field dynamically adjusts to a smooth
equilibrium.
Regions of high $Q$ are found to be better related to regions of high integrated parallel 
 current than to actual current sheets.

\end{abstract}

{\bf Keywords:}  magnetohydrodynamics (MHD); magnetic fields; solar corona.

\vfill

\begin{center}
$\star$ corresponding author: antonia@maths.dundee.ac.uk
\end{center}

\newpage

\section{ Introduction}

Several different mechanisms have been proposed as explanations
for solar coronal heating and indeed it is likely that a number of different
heating processes are responsible for the extreme temperature values
found.   Much of the corona is considered to be close to force-free, that is, in
 a state satisfying    $\mathbf{j} \times \mathbf{B} \simeq \mathbf{0}$.
This consideration led to the proposal of one of the earliest and most debated coronal
heating theories, that of topological dissipation (Parker, 1972). Parker theorized
that the coronal magnetic field is unable to relax to a smooth force-free equilibrium 
following arbitrary perturbations of the field via footpoint motions and 
the consequence is a development of tangential discontinuities in the field.
 Magnetic reconnection may then occur across the resulting current sheets.
 Since Parker's original paper a multitude of arguments for and against the  
 hypothesis have been given but no unanimity presently exists
(e.g.~van Ballegooijen, 1985; Galsgaard \& Nordlund 1996, Craid \& Sneyd 2005).

One technique presently employed for predicting current sheet formation in 
continuous fields  is to examine
a quantity known as the squashing factor, $Q$ (Titov {\it et al.} 2002; see Appendix A for 
further details of the function itself).
This is a measure for continuous magnetic fields which describes the level of 
squashing of an infinitesimal  flux tube; $Q$ is large in regions where the footpoint 
mapping of the field is highly distorted.  Regions of very high $Q$ outline so-called 
quasi-separatrix layers or QSLs (Priest \& D{\'e}moulin 1995) and 
can be considered as an analogue of the separatrix surface in configurations with no null points.
 QSLs are thought to be sites of preferential current sheet formation
 (D{\'e}moulin {\it et al.} 1996; Titov {\it et al.} 2002)
  and a number of numerical experiments have supported this hypothesis
 (e.g.~Longcope \& Strauss 1994; Milano {\it et al.}~1999;  Galsgaard {\it et al.} 2003;
 Aulanier {\it et al.}~2005).  In addition, observational studies have attempted to examine 
 the relation between QSLs and current sheets (e.g. Schmieder {\it et al.} 1997; 
 Fletcher {\it et al.} 2001) although this is difficult due to the spatial resolution and quality of 
the data required for an accurate depiction of the QSLs.

 In an earlier paper (Wilmot-Smith {\it et al}.~2009) we considered a class of braided
magnetic fields.  The braids were initially given in closed form (and not in MHD equilibrium),
then inserted into a numerical scheme and underwent magnetic relaxation towards a force-free
equilibrium.  In that paper the current structure of the fields was looked at in some detail.  The
results are summarized for convenience and put into context in Section~3.
Here we return to these models for braided fields in order to examine the relationship
between QSLs, currents and a quantity which is important for magnetic reconnection in 
three dimensions,  the electric current integrated along magnetic field lines.
As such we begin by describing the braided fields under consideration.

 \section{Model for a class of braided fields}%, $E^{n}$}
 
\begin{figure}[ht]
\begin{center}
\subfigure[]{\label{fig:edge-0a}\includegraphics[width=0.25\textwidth]{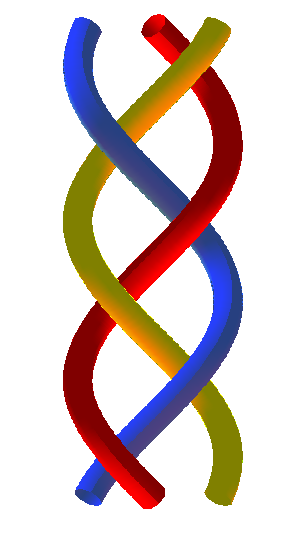}}
\subfigure[]{\label{fig:edge-0b}\includegraphics[width=0.2\textwidth]{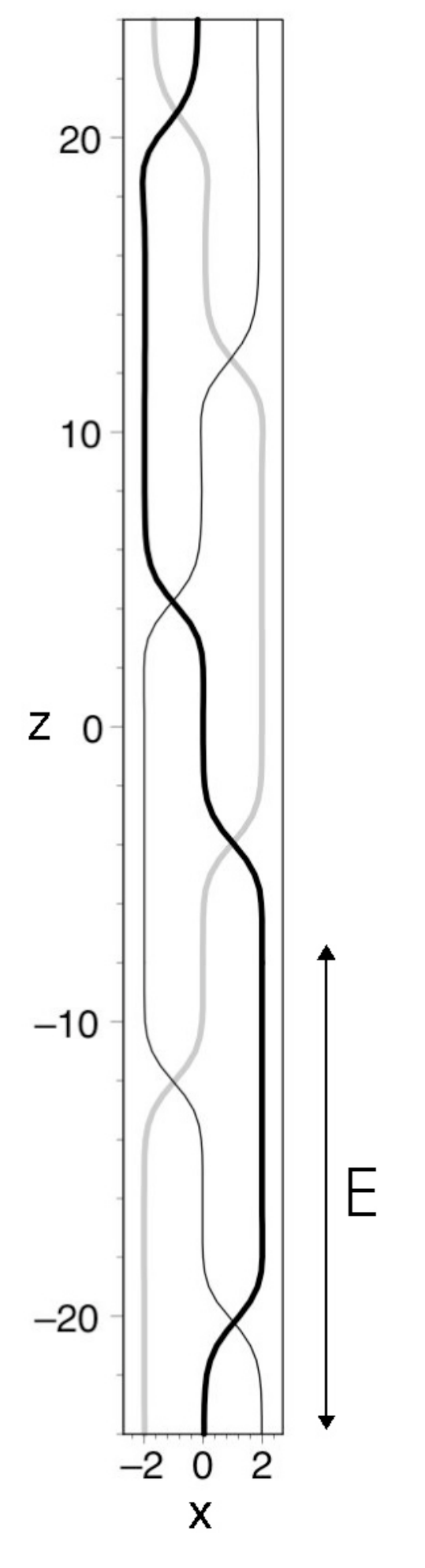}}
\end{center}
\caption{(a) The pigtail braid and (b) some field lines from
the field $E^{3}$ modelled on the pigtail braid.
$E^{3}$ may be considered a concatenation of three elementary parts, $E$ (indicated
in the figure) and,
accordingly, $E^{n}$ may be constructed for various $n$ via similar concatenations.}
\label{fig:100}
\end{figure}

 In Wilmot-Smith {\it et al}.~2009, (hereafter Paper I) we introduced a class of braided
 fields designed to model magnetic fields in solar coronal loops.
 The fields are given as  $E^{n}$, $n \in \mathbb{Z}^{+}$,
where $E^{n}$ can be considered as a concatenation of $n$ times the `elementary' field $E$.
 The braided field $E^{3}$ is modelled on the pigtail braid as illustrated in Figure~\ref{fig:100}.

 The nature of $E$ and the construction mechanism for $E^{n}$ may be seen from
that figure.  The field $E$ consists of two localized regions of twist in an otherwise 
uniform field. These two regions are of opposite sign but are of the same underlying form -- 
the superposition onto a uniform vertical field, $b_{0} \hat{\bf z}$, of a toroidal flux ring given by
\begin{equation}
\mathbf{B}_{c} = 2 b_{0} k \frac{r}{a} \textrm{exp}  
                               \left( -\frac{r^{2}}{a^{2}} - \frac{z^{2}}{l^{2}} \right) \mathbf{e}_{\phi}. \notag
\end{equation}
In Cartesian coordinates, placing the centre of this region of toroidal flux at the point 
$(x_{c},y_{c},z_{c})$ we have
\begin{equation}
\label{eq:unittwist}
\mathbf{B}_{c} = 
2  \frac{b_{0} k}{a}  \textrm{exp}\left(
\frac{ \scriptstyle-\left(x-x_{c}\right)^{2}-\left(y-y_{c}\right)^{2}}
{ \scriptstyle a^{2}} - \frac{ \scriptstyle \left(z-z_{c}\right)^{2}}{\scriptstyle l^{2}}  \right)
\left( -  \left(y - y_{c} \right) \mathbf{\hat{x}}  + \left(x - x_{c} \right) \mathbf{\hat{y}}  \right),
\end{equation}
where $c=\lbrace x_{c},y_{c},z_{c},k,a,l \rbrace$ and $b_{0}=1$ is taken throughout.
 The closed form expression  for $E$ is given by  
\begin{equation}
\label{eq:2blobbbbs}
b_{0} \hat{\bf z} + \sum_{i=1}^{2} \mathbf{B}_{c_{i}} \notag
\end{equation}
where
$c_{1}=\lbrace  1,0,-4,1,\sqrt{2},2 \rbrace$  and $c_{2}=\lbrace -1,0,4,-1,\sqrt{2},2 \rbrace$
and, similarly, that for $E^{n}$ by  
\begin{equation}
b_{0} \hat{\bf z} + \sum_{i=1}^{2n} \mathbf{B}_{c_{i}} \notag
\end{equation}
with appropriate choices for the parameters $c_{i}$.  For each field 
$E^{n}$ we are able to find the equations of the magnetic field lines in closed form 
and hence an exact expression for $Q$ for each $n$ can be easily calculated.

\section{Current structures}  %in the braided fields}% $E^{n}$}
 
In paper I a Lagrangian numerical scheme (Craig \& Sneyd, 1986) was used
to carry out an ideal relaxation of  $E^{3}$ towards a force-free state with the field on 
the boundaries of the domain fixed. The maximum Lorentz force within the domain 
decreased by two orders of magnitude during the relaxation and the resultant field 
is  close to force-free (for a precise definition of what is meant by `close' see Section~3.2 of Paper I).  
Beyond this point problems with the numerical accuracy of the scheme (Pontin {\it et al.}, 2009) 
mean no further relaxation is possible until the numerical difficulties are resolved.
In the relaxed state for $E^{3}$ the current is  smooth and of a large-scale (see Figure~3 of
Paper I) -- two regions of current extend vertically throughout the domain and the maximum
magnitude of the current, (${\bf j}_{max}  = 1.47$), is around half of that of the initial state.  
No evidence for current sheet formation in the ideal relaxation was found.
 
 It is well known that continuous motions on the boundary of an initially smooth magnetic field cannot 
 lead to truely {\it singular} current sheets under an ideal (line--tied) evolution (Van Ballegooijen 1985).
 However it might seem that the lack of current sheets in our relaxed state contradicts previous results,
 for example those of Longbottom {\it et al.}~(1998)  and Galsgaard {\it et al.}~(2003).  These investigations 
 found the formation current concentrations  under shearing motions on the boundary of an initially 
 current-free configuration.  Since we can think of our braid as having been created by a sequence of such shearing 
 motions applied to  an initially homogenous field, we might also expect to find similar current concentrations 
 in the relaxed state. A closer comparison, however, shows that the twist, measured  by the ratio of the displacement 
 distance  ($l_d$) to the length of the domain ($L$) in which current concentration are observed is comparatively 
 high in the experiments of Longbottom {\it et al.}~(1998)  and Galsgaard {\it et al} (2003).
  In particular,  Longbottom {\it et al.}~(1998) looked in detail into this question and found significant build-up of 
 current for values above  $l_d/L \approx 0.6$ ($l_d$= `shear amplitude' in their terminology and $L=1$). 
 If we were to create our elementary braid $E$ (which comes closest to the configuration they considered)  
 by a similar shearing pattern, we would obtain a ratio  $l_d/L < 0.18$ ($l_d \approx 2 \sqrt{2}$; $L=16$). 
 This is a value for which no noticeable current concentration has been observed (see Figure 2 of 
 Longbottom {\it et al.}~1998) in agreement with our findings. 
 A similar argument applies to the conditions under which Galsgaard {\it et al.}~(2003) found current concentrations, although their experiment does not allow a direct comparison as they used a resistive code. In their case, values of  
$l_d$ ( `shear distance' in their terminology, $L=1$) below $0.2$ also showed no significant build up of current.

 While this shows that our results do not contradict previous findings of current sheet formation, 
 it is perhaps still surprising that a smooth near force-free equilibrium for such a complex configuration exists. 
  In order to understand this one has to bear in mind that our equilibrium is not an exact force-free state but only close to a 
 force-free state (for an exact definition how close see Section 3.2 of paper I).   If the residual forces in the relaxed 
 state could be balanced by a pressure, our relaxed state would have a plasma beta of $ \beta \approx 0.009$; 
 although this is well within the  quality of `force-freeness' usually assumed for magnetic loops in the solar corona, 
 it is not zero. Moreover, the field has comparatively low maximum force-free field  parameter $\alpha$ (when measured
 in a dimensionless way) and it is known that for sufficiently small values of this quantity, force-free equilibria exist 
 for arbitrarily braided fields (Bineau, 1972). Unfortunately the proof given does not give an explicit upper bound but 
 only provides the existence of such a number. These two properties, a low $\alpha$ and an only approximately 
 force-free state,  appear to allow for a sufficiently large space of smooth solutions to encompass the braids we have
  investigated.
 
 %%% ???
  Whilst in two-dimensions the current itself is the crucial quantity for 
  magnetic reconnection, in three-dimensions 
 the integrated parallel electric field along magnetic field lines plays an important role (Hesse \& Schindler, 1988).  In
resistive MHD, the parallel electric field is related to the parallel current by the relation
 $\int E_{\parallel} dl =  \eta \int j_{\parallel} dl$ (assuming a uniform resistivity $\eta$).
Motivated by this consideration, in Paper I we  examined the integrated parallel current,  
hereafter denoted by $\mathcal{J}_{\parallel}$,  in both the initial and relaxed states for 
$E^{3}$.  In both cases a highly filamentary structure containing multiple thin $\mathcal{J}_{\parallel}$ layers was 
found.   During the relaxation the spatial structure of $\mathcal{J}_{\parallel}$ was approximately conserved. 
More generally, by considering $E^{n}$ for various $n$, the  thickness, $d$, of the $\mathcal{J}_{\parallel}$ layers 
was found to decrease exponentially with $n$ (degree of braiding). 
 %%%

 A quantity proposed as an indicator for the formation of current sheets (and hence for the occurrence of 
 magnetic reconnection)  in continuous fields is known as the squashing factor, $Q$ (Titov {\it et al.}~2002; 
 Titov {\it et al.}~2009).   Layers in which $Q$ is large, so-called quasi-separatrix layers  (QSL),  can be understood 
 as a generalisation of separatrix surfaces for cases where no null points exist in the domain under consideration.
 Our previous findings motivate a consideration of the nature of $Q$ in the class of braided magnetic fields
 $E^{n}$.   We relate these findings both to the current structure and the integrated parallel current structure
 of the fields.
  
\section{Squashing factor and QSLs}
 
 A method for obtaining the squashing factor $Q$ in the braided fields considered is outlined
 Appendix A.  Note  that for $E^{n}$ the squashing factor $Q$ and the newer definition of the slip-squashing 
factor (Titov 2007, Titov {\it et al.} 2009) give the same result since the magnetic field is 
normal to both $z$-boundaries.   
 
 We begin the discussion by considering the simplest field, $E$.  Under ideal relaxation towards 
a force-free equilibrium, the current, ${\bf j}$, in $E$ remains smooth and large-scale,
with the maximum value being ${\bf j}_{max} = 1.07$ in the relaxed state
(compared with ${\bf j}_{max} = 2.83$ initially).
\begin{figure}
\begin{center}
\includegraphics[width=\textwidth]{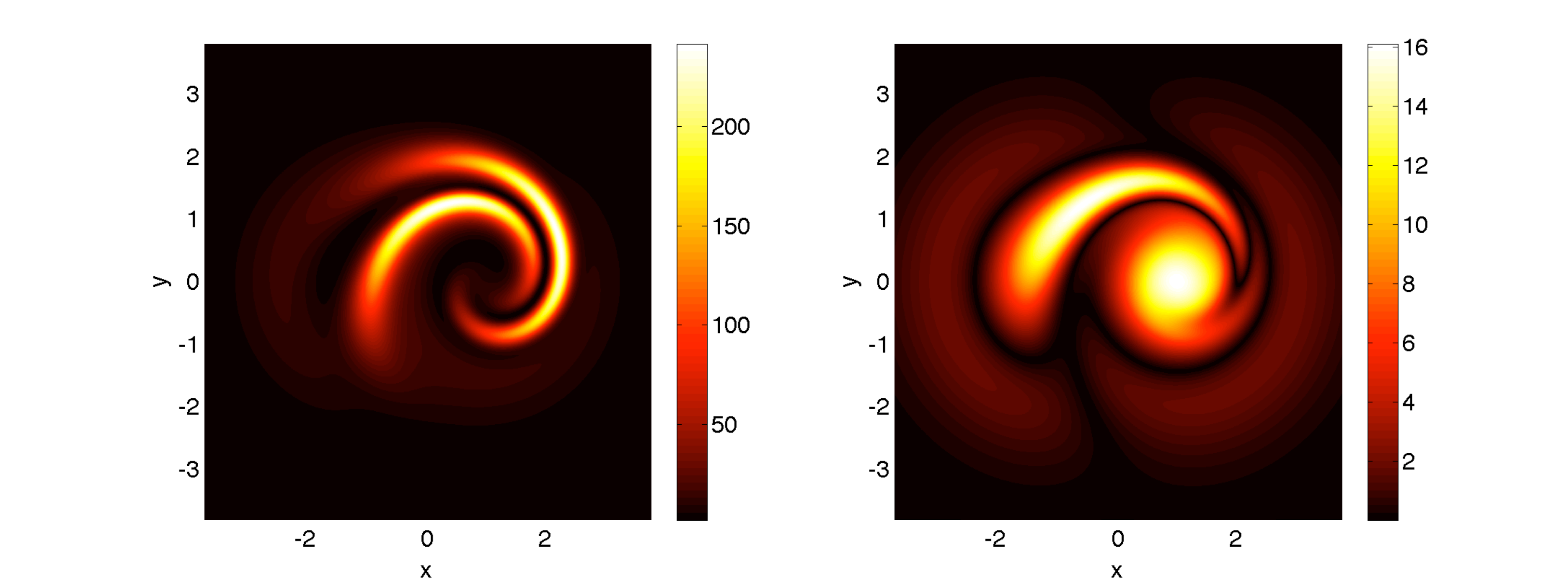}
\end{center} 
\caption{({\it left}) $Q$ as a function of field line positions on the lower 
boundary of $E$ and ({\it right}) integrated parallel current along field lines,
$\vert \mathcal{J}_{\parallel}\vert$, on the same boundary
for the field $E$. \label{fig:qje}} \end{figure}
The squashing factor $Q$ for $E$ is shown in the left-hand image of
Figure~\ref{fig:qje} as a function of field line positions on the lower boundary.
High values of $Q$  (regions where the field line mapping is strongly distorted)
are found in two layer-like regions, and the maximum value of $Q$ is $Q_{1,max} = 241.5$.
Contours of the parallel current integrated along field lines, $\mathcal{J}_{\parallel}$,
 are shown in the right-hand image of Figure~\ref{fig:qje}.
  The characteristic scales of  $\mathcal{J}_{\parallel}$ are larger than those of 
 $Q$ and no relation between the two is evident.
 
 \begin{figure}[ht]
\begin{center}
\includegraphics[width=0.49\textwidth]{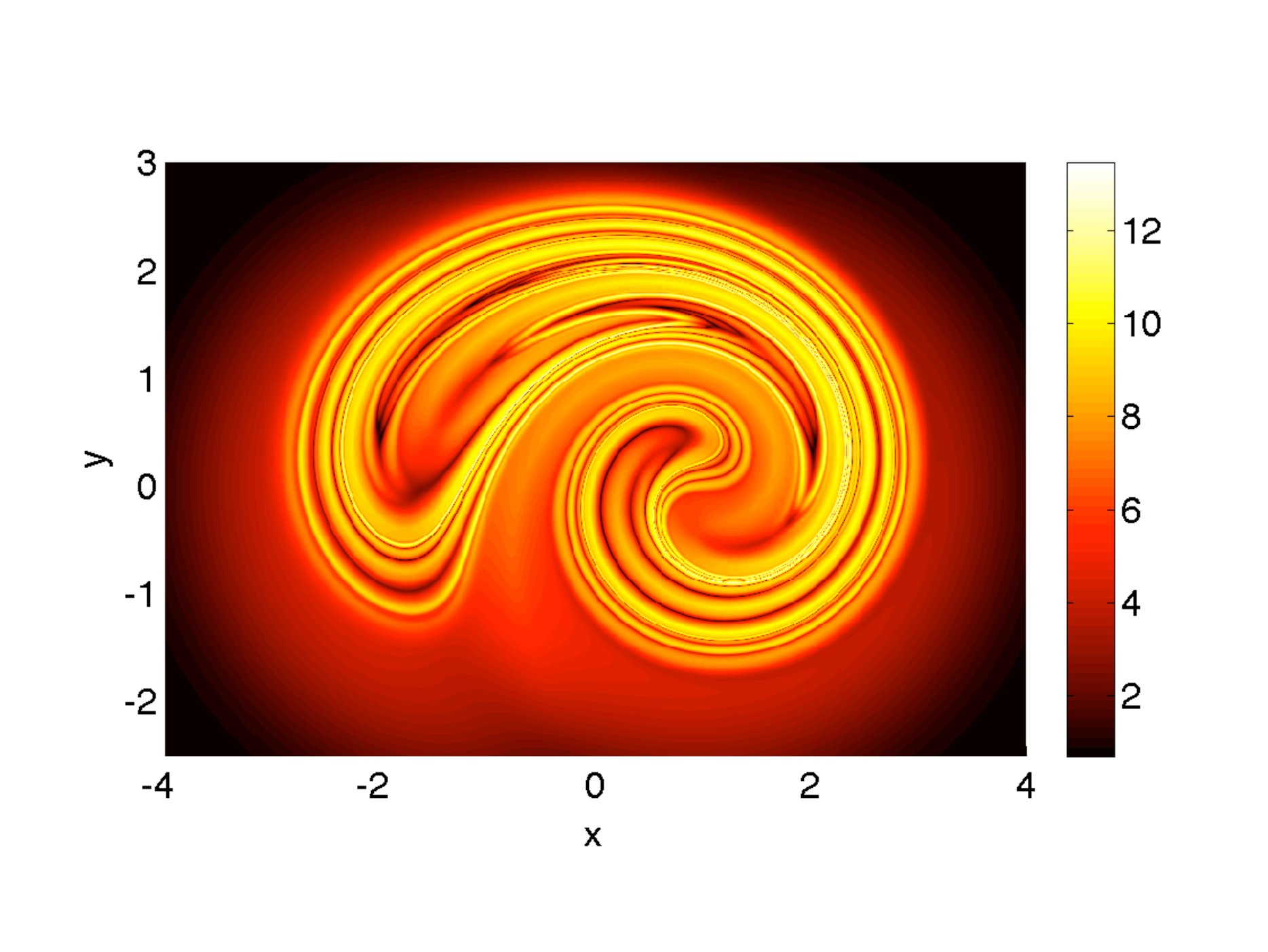}
\includegraphics[width=0.49\textwidth]{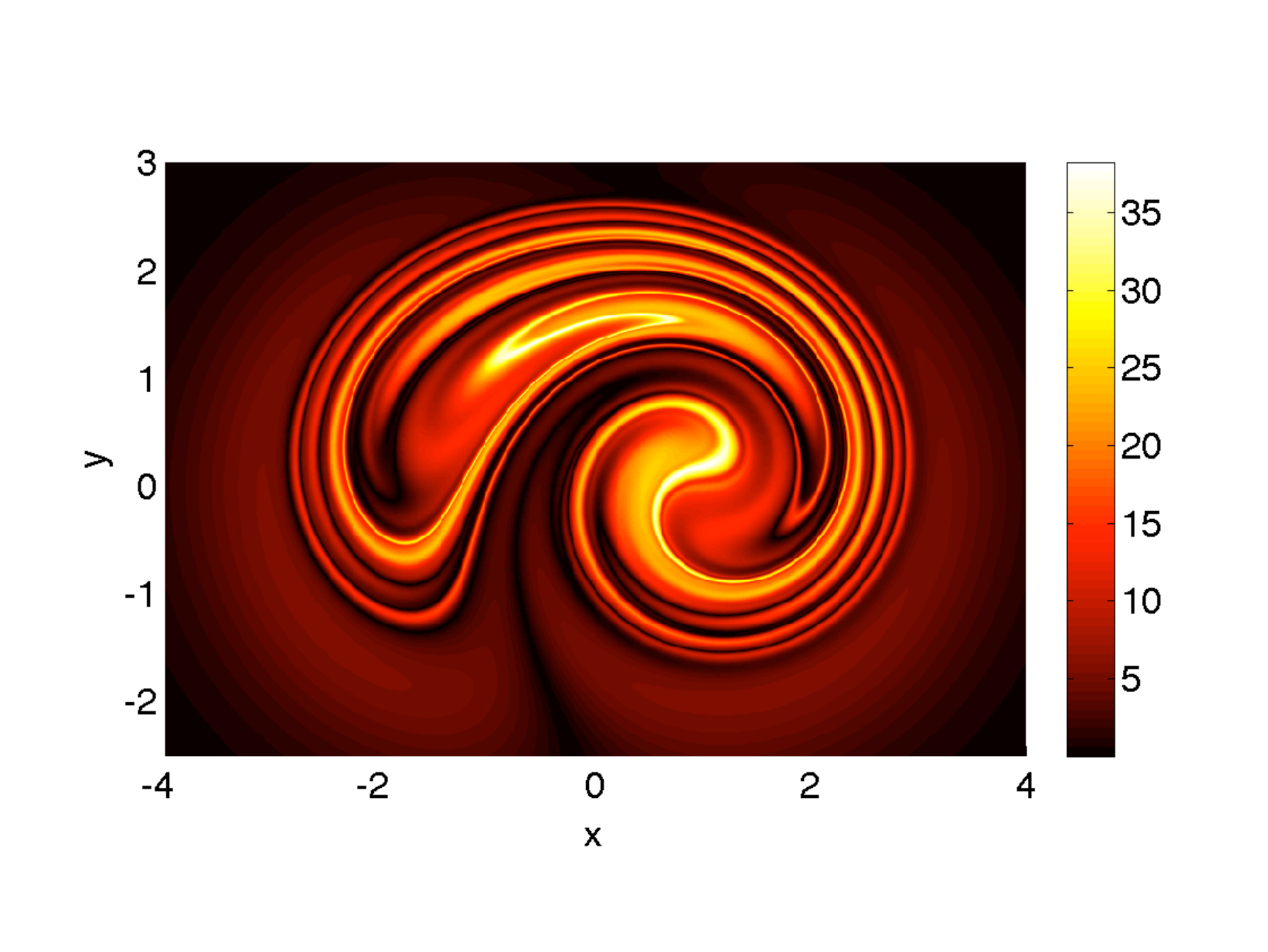}
\end{center}
\caption{({\it left})  $\log_{10}(Q)$  as a function of field line position on the
lower boundary and  ({\it right}) integrated parallel current along field lines,
$\vert \mathcal{J}_{\parallel}\vert $, on the same boundary
for the braided field $E^{3}$ (based on the pigtail braid).}
\label{fig:0}
\end{figure}
Next consider the field $E^{3}$, that based on the pigtail braid and on which most of the analysis of Paper I was 
based.  Recall that a smooth equilibrium with only large-scale current structures was obtained 
in an ideal relaxation of this field.  
 
 The squashing factor $Q$ on the lower boundary of the domain is shown in the 
left-hand image of Figure~\ref{fig:0}.
  Here the maximum value of $Q$ is  $Q_{3,max}=7.17 \times 10^{5}$ 
 (with $Q$ calculated at $1396^{2}$ points on the lower boundary) and
  regions of enhanced $Q$ occur in a multitude of thin layers in a large portion of the domain. 
Between these layers $Q$ drops significantly, even to its minimum possible value of $2$ in many 
locations. A typical half-width at half-maximum (HWHM) 
of the QSLs is $\sim 10^{-3}$ or  $0.01$\% of the domain width.

\begin{figure}
\begin{center}
\includegraphics[width=1.08\textwidth]{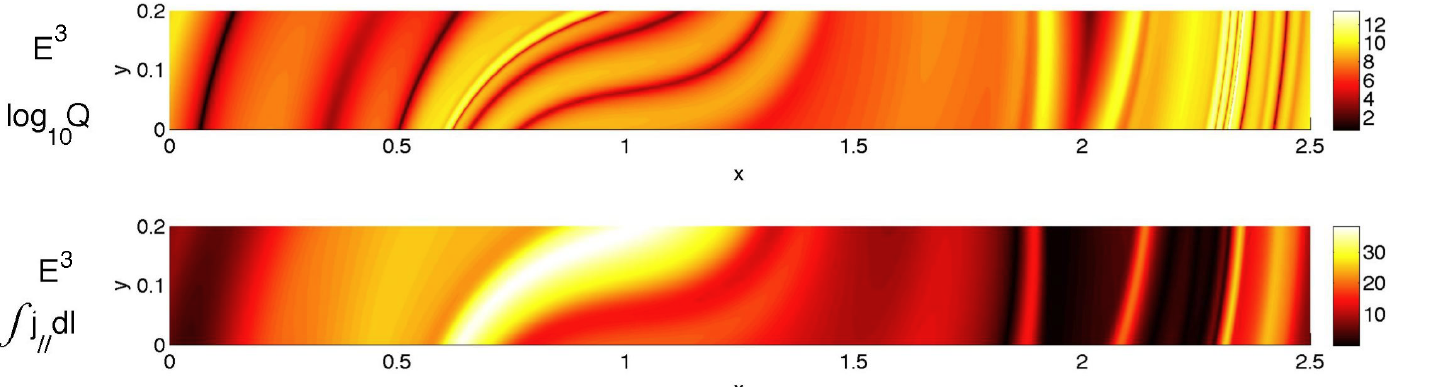}

\vspace*{1.5cm}
\includegraphics[width=1.08\textwidth]{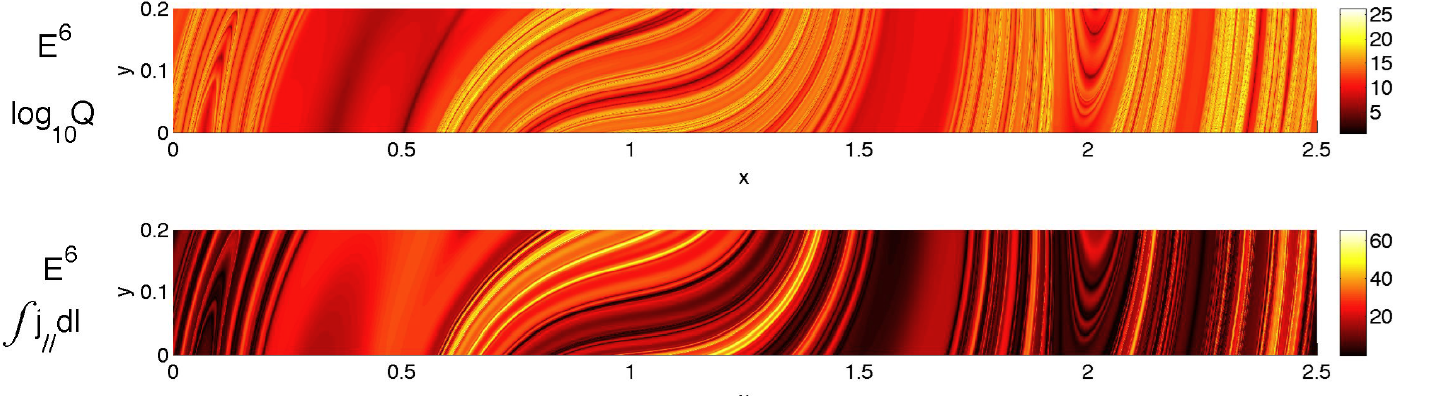}
\end{center}
\caption{Figure showing
 $\log_{10}(Q)$ and the absolute 
integrated parallel current along field lines $\mathcal{J}_{\parallel}$ for
$E^{3}$ (top panel) and $E^{6}$ (lower panel).
  For clarity, only a subsection of the domain has been shown. }
\label{fig:1}
\end{figure}

The integrated parallel current structure, $ \mathcal{J}_{\parallel} $, for $E^{3}$
is shown on the lower boundary of $E^{3}$ in the right-hand image of
Figure~\ref{fig:0}. A resemblance between $Q$ and 
$ \mathcal{J}_{\parallel} $ is shown, with both quantities having a similar global 
structure and containing thin layers of approximately equal width. However,
$Q$ is more filamentary than $ \mathcal{J}_{\parallel} $ and has a much greater range of values.
In order to make a more precise comparison, the reader is referred to the upper panel
of Figure~\ref{fig:1} where both quantities are shown for only a subsection of the domain.

Finally consider the field $E^{6}$.  Both $Q$ and $ \mathcal{J}_{\parallel} $ 
are shown over a subsection of the domain in the lower panel of Figure~\ref{fig:1}.
The structures of the two quantities are highly filamentary, both containing 
extremely small scales (HWHM of $\sim 10^{-5}$). 
It is seen that $Q$ and $ \mathcal{J}_{\parallel} $ here have a strong resemblance
in terms of the global pattern of their thin layer-like structures.  However, the locations
of the regions of high (low) $Q$ and high (low) $ \mathcal{J}_{\parallel}$ do not 
exactly coincide.

\begin{figure}
\begin{center}
\subfigure[]{\label{fig:edge-1a}\includegraphics[width=0.49\textwidth]{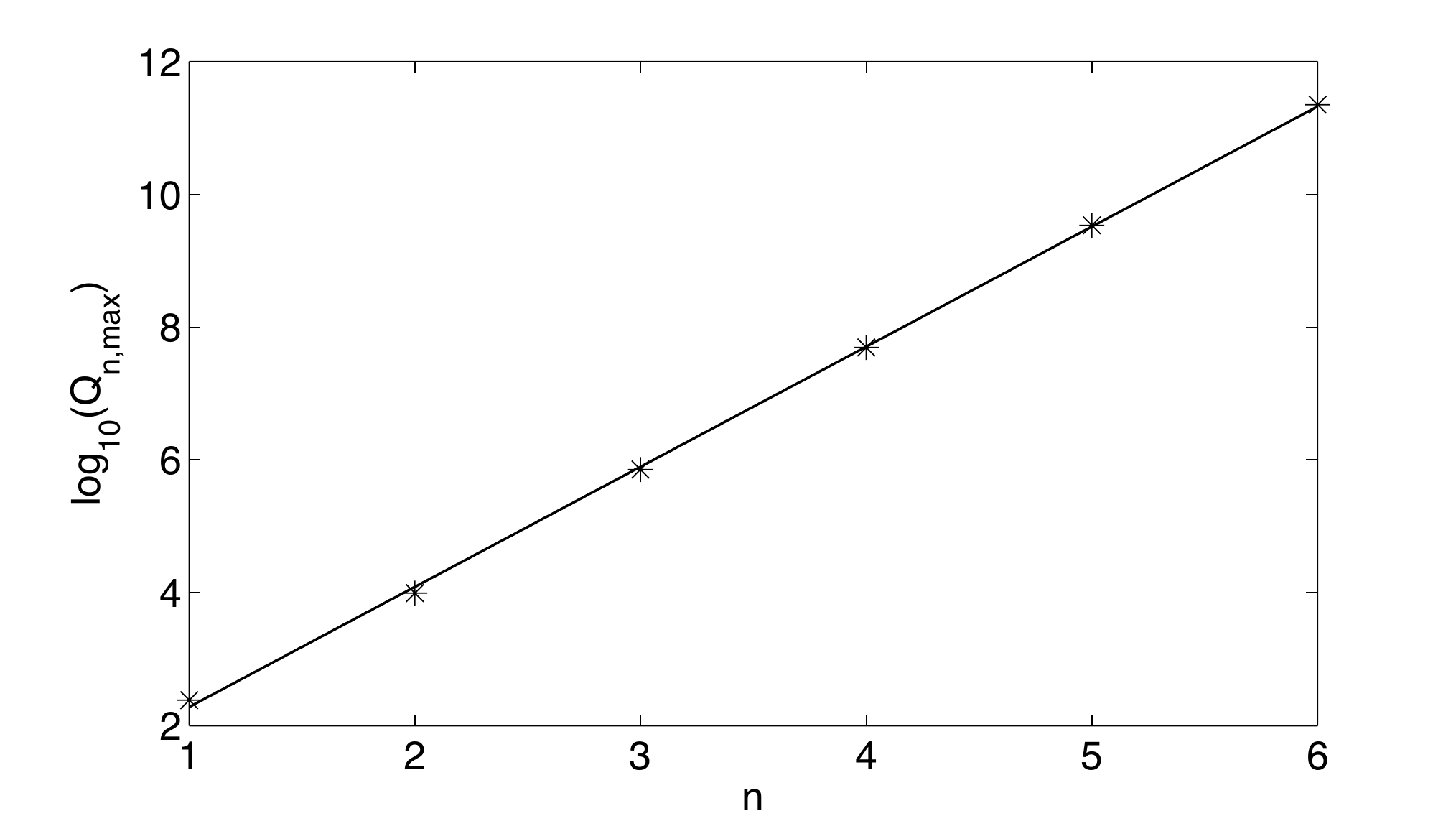}}
\subfigure[]{\label{fig:edge-1b}\includegraphics[width=0.49\textwidth]{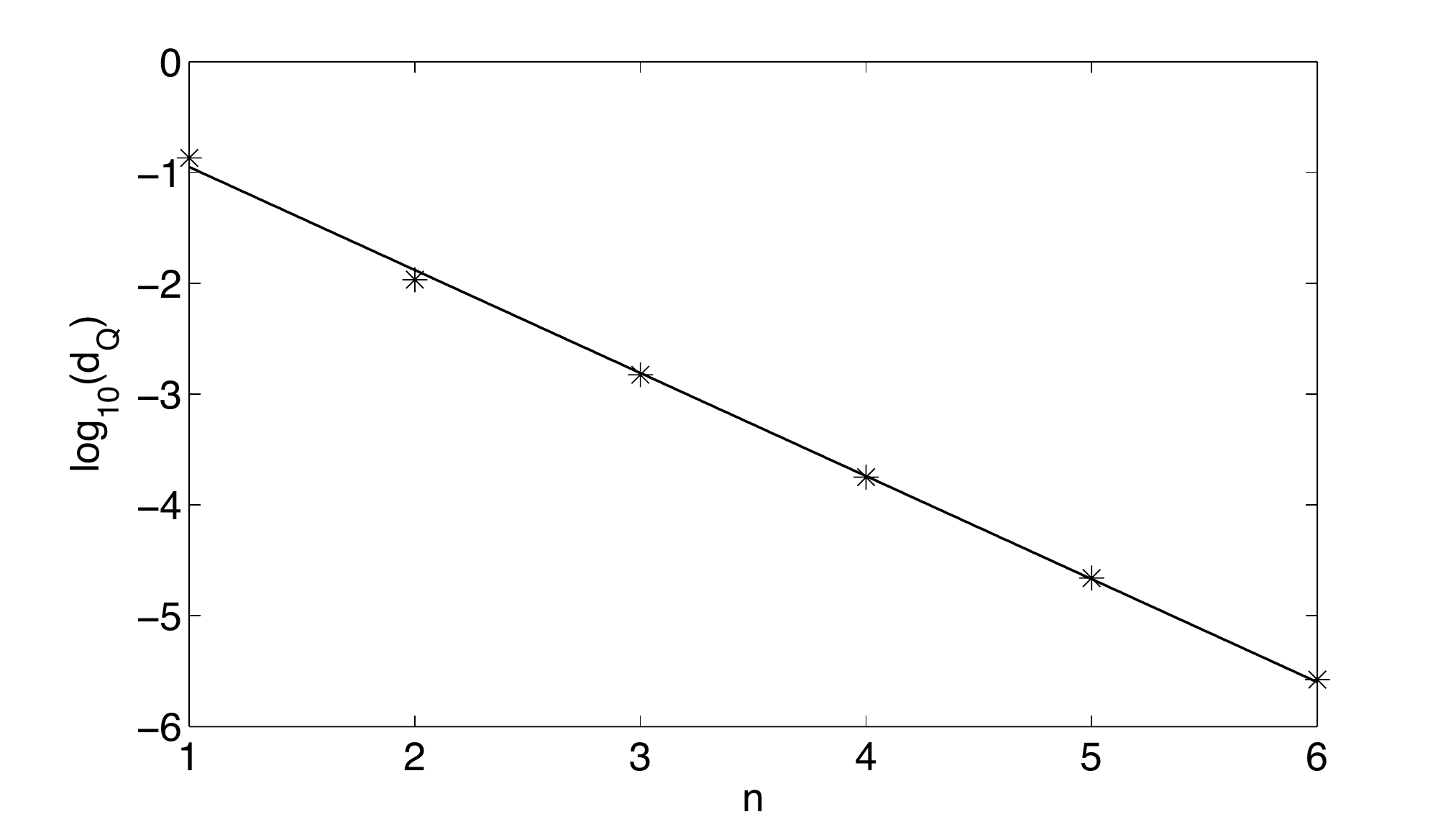}}
\end{center} 
\caption{(a) Maximum value of $Q$ with $n$ in the full domain.  An exponential  
exponential increase in $Q$ with $n$ is shown.
(b) HWHM of $Q$ layers with $n$.
An exponential decrease in $Q$ with $n$ is found.
Here the width has been taken as the shortest such width along a cross section 
of the domain on the lower boundary. }
\label{fig:2}
\end{figure}

Considering now $Q$ alone for the various fields $E^{n}$, 
we seek to determine how the maximum value of $Q$
and the characteristic width (specifically HWHM), $d_{Q}$, of the layers of high $Q$ depend on $n$.
It is found that the maximum of $Q$ for $E^{n}$, $Q_{n,max}$, in the domain increases 
exponentially with $n$ -- as shown in  Figure~\ref{fig:2}(a) -- for which a best fit line is
\[ Q_{n,max} = 2.97 \times 10^{1.8n}.\]
In Paper I a linear increase in the maximum of $\mathcal{J}_{\parallel}$ with $n$ was found,
as expected since the vertical length of the domain is also increasing linearly
by construction of the fields.
The typical width, $d_{Q}$, of the regions of enhanced $Q$
 decreases with $n$.  More precisely, there is an exponential decrease
in $d_{Q}$  with $n$ -- as demonstrated in Figure~\ref{fig:2}(b) -- for which 
a best fit line is $$d_{Q} = 1.90 \times 10^{-0.93n}.$$
In Paper I a similar exponential decrease in the width ($d_{\mathcal{J}_{\parallel}}$)
of regions of high $\mathcal{J}_{\parallel}$ was found.

We now  address the dependence of the maximum value of $Q$, 
$Q_{n,max}$ on $n$.  Recall that $Q_{n,max}$  was found to increase exponentially
with $n$.
Notice that by the construction of $E^{n}$ we may place an upper bound on the maximum 
value of $Q$ from the method of its calculation by composition of matrices.
Letting $Q_{i}$ denote the maximum value of $Q$ for $E^{i}$, then
the submultiplicative property  of the $p=2$ norm tells us that
$Q_{n,max} \leq \left(Q_{1,max}\right)^{n}.$  
Evaluating, we obtain the upper bound as
$\left(Q_{1,max}\right)^{n} \sim 10^{4.7n}/200,$
whilst the true dependence is given by
$ Q_{n,max} \sim 2.97 \times 10^{1.8n}.$
The dependence on the smaller power then comes from the filling factor of the field $E$,
i.e.  the number of points on the lower boundary of $E$ with highest $Q_{1}$
that get mapped under $\bf{F}$ to points with high $Q_{1}$ 
for the next composition (for $E^{2}$ etc.)
 
In the next section we seek to explain these findings and explore theoretically
the relationship between $Q$ and the current structures.

 \section{Relating QSLs and current structures.}

 As mentioned above, the squashing factor has been suggested as an indicator for the
 formation of current sheets.  Layers in which $Q$ is large outline quasi-separatrix layers
 (QSLs).  The link between QSLs and current sheets relies on the following hypothesis (referred to below as 
the QSL--hypothesis).   Where $Q$ is high there is a sensitive dependence of the end points of magnetic field 
lines on the starting points (or vice-versa since $Q$ is symmetric). 
Hence certain motions of the plasma on one boundary (particularly those crossing the QSLs)
 would lead to very high velocities at the other boundary, provided the evolution is ideal 
and the magnetic configuration is unaffected by this perturbation. For sufficiently high $Q$ this is inconsistent 
with ${\bf v} < {\bf v}_A$ and so the magnetic configuration has to change. Then a lack of neighbouring smooth 
equilibria can lead to the formation of current concentrations or even current sheets.

 Within the experiments described above we are in a position to test the QSL--hypothesis, given that we 
 can think of the relaxed configuration for $E^{3}$ as as an ideal deformation of the initial configuration.
The initial configuration 
  shows an abundance QSLs but  no strong current concentrations. 
  In the subsequent ideal relaxation the QSLs remain but no current sheets (in the sense of a sheet-like structure 
 of  ${\bf j}$)  form. This result appears to contradict the QSL--hypothesis.   There are, however, two ways in which 
 the system can escape the necessity of forming current concentrations. 

 Firstly, the deformation applied could be unfavourable for the formation of current sheets
 (Titov {\it et al.}~2003).  For example, Galsgaard {\it et al.}~(2003) considered a configuration containing
 two intersection QSLs and applied two types of deformations on the boundary. They found 
 that only one of the types of deformation lead
  to significant build-up of current.  
  
  Given the complexity of our configuration, it seems unlikely that all of the motions which
  occur during the relaxation are of the type unfavorable for the formation of current sheets 
 unless the nature of the deformation applied (${\bf v} \sim {\bf j} \times {\bf B}$) 
excludes any such favorable deformation.
  The second possibility is that following the applied deformations the system is able to adjust to a 
 neighbouring smooth equilibrium.  This possibility is supported by the arguments given in Section~3 that 
 our final state is a low alpha, near force-free state.  

\begin{figure}
\begin{center}
\includegraphics[width=1.0\textwidth]{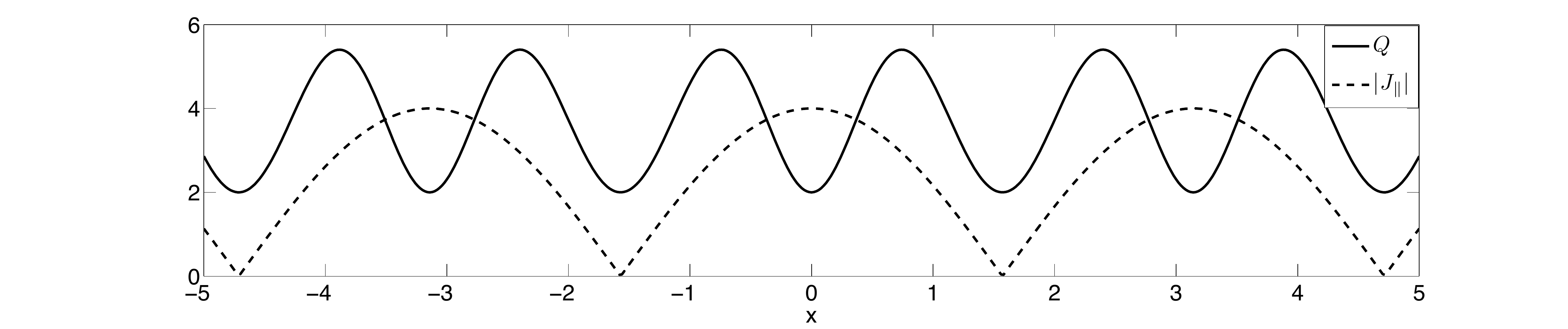}
\end{center}
\caption{$Q$ (solid line) and $\vert \mathcal{J}_{\parallel}\vert $ (dashed line)
for the illustrative example of a particular 1D force-free field (described
in Section~5).   Here $\vert \mathcal{J}_{\parallel}\vert = 4 \vert {\bf j}\vert$.
In this example $Q$ is not a good predictor of current features.
(Parameters: $c_{1}=0$, $c_{2}=1$, $\lambda=1$, $s=4$.)}
\label{fig:exQJ}
\end{figure}

Although current layers are not found in our experiment, it does show layer-like structures of $\mathcal{J}_\|$. 
 Such layers have a broader definition in that only the integral of ${\bf j}$ and not  ${\bf j}$ 
itself needs to show small scales.  The following  example, however, shows that 
there is no simple relation between either $Q$ and ${\bf j}$ or between  $Q$ and $\mathcal{J}_\|$. 
We have chosen this example because it is locally similar to the braided fields discussed, in that 
in both cases $Q$ and  $\mathcal{J}_{\parallel}$ each have a layer-like structure,
with multiple layers in which the quantities are enhanced. Consider the force-free magnetic field
\[ \mathbf{B} = \left( c_{1} \sin \left( \sin\left(\lambda x\right)/\lambda\right)
+c_{2}  \cos \left( \sin\left(\lambda x\right)/\lambda\right) \right)  \hat{\bf y}
+ \left( c_{1} \cos \left( \sin\left(\lambda x\right)/\lambda\right)
-c_{2}  \sin \left( \sin\left(\lambda x\right)/\lambda\right) \right)  \hat{\bf z} \]
for which $ \mathbf{j} = \alpha \mathbf{B} $ with $ \alpha = \cos \lambda x $,
and take  $z \in [0,s]$, $x,y,\in (-\infty,\infty)$.
On the lower boundary, $z=0$, we have
\[Q = 2 + s^{2} \cos^{2}(\lambda x) \left(c_{1} \cos \left( \sin\left(\lambda x\right)/\lambda\right)
-c_{2}  \sin \left( \sin\left(\lambda x\right)/\lambda\right) \right)^{2}, \]
\[ \mathcal{J}_{\parallel} = \int_{s'=0}^{s} {\bf j} \cdot {\bf B} \ ds' = 
s (c_{1}^{2}+c_{2}^{2}) \cos \lambda x, \]
and
\[ \vert {\bf j} \vert^{2} = (c_{1}^{2}+c_{2}^{2})  \cos^{2}(\lambda x).\]
Here the relationship between $Q$ and currents crucially depends on the
choice of parameters.
Setting $c_{2}=0$, for example, maxima and minima of  $Q$, 
$\vert \mathcal{J}_{\parallel} \vert$ and $\vert {\bf j}\vert$ coincide.
Setting instead $c_{1}=0$, then minima in $Q$ correspond to both
maximal and minimal regions of $\vert \mathcal{J}_{\parallel}\vert$ or $|{\bf j}|$
(see Figure~\ref{fig:exQJ}).
Here we see that the maxima of $Q$ increase quadratically with the 
vertical extent ($s$) of the domain under consideration, whilst
the maxima of $\mathcal{J}_{\parallel}$ increase only linearly and the maxima of $|{\bf j}|$ are 
independent of $s$. 

Note that this is a specific illustration of a more general case of a force-free
field $\mathbf{B} = B_{y}(x) \hat{\mathbf{y}} + B_{z}(x) \hat{\mathbf{z}}$
with force-free parameter $\alpha(x)$. Taking $z \in[0,s]$ then
on the lower boundary $Q=2 + s^{2} \alpha^{2} B_{z}^{2}$ while 
$ {\bf j}  =  \alpha \left(B_{y}^2 + B_{z}^2\right)^{1/2}$  and
$\int j_{\parallel} dl  = s \alpha \left(B_{y}^2 + B_{z}^2\right)^{1/2}$ .

The above described situation corresponds well to the braided fields $E^{n}$ in 
that both $Q$ and $\mathcal{J}_{\parallel}$ have a layer-like structure.  
For $E^{n}$ we also find that there is no exact correspondence between the layers of 
high $Q$  and greatest $\mathcal{J}_{\parallel}$.
However, there is a similarity between $Q$ 
and $\mathcal{J}_{\parallel}$ in terms of their global structure, particularly 
for moderate and large $n$ (see Figures \ref{fig:0} and \ref{fig:1}).
This similarity comes not from an inherent relation between $Q$ and $\mathcal{J}_{\parallel}$
but rather from the fact that both quantities are functions of the field lines.

 \begin{figure}
\begin{center}
\includegraphics[width=1.0\textwidth]{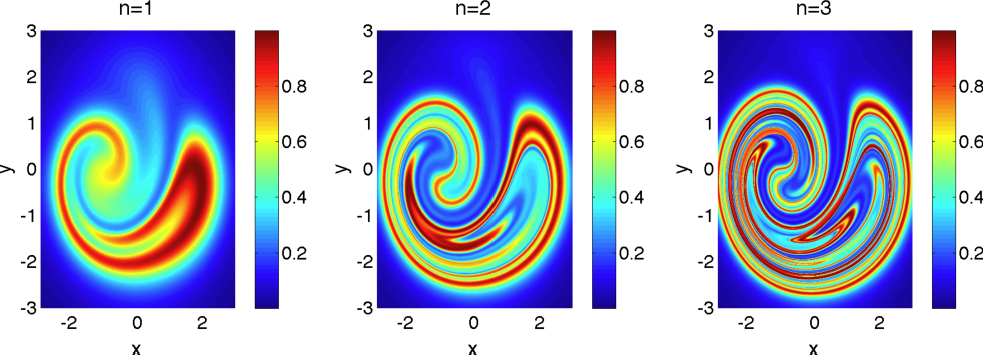}
\end{center}
\caption{Illustrative example demonstrating properties of the 
field line mapping ${\bf F}^{n}(x,y)$.
  For the function  $\psi = \exp(-x^{2}/4-y^{2}/4)$, the left hand image 
 shows   $\psi({\bf F}(x,y))$ 
 centre image shows $\psi({\bf F}^{2}(x,y))$ 
and the right hand image shows $\psi({\bf F}^{3}(x,y))$.
As ${\bf F}$ is repeatedly applied to the function the image develops
small scales;  for this reason since $Q$ and $\mathcal{J}_{\parallel}$ 
take (by definition) one value on each field line, they too must develop 
small scales on the boundaries of $E^{n}$ for increasing $n$.}
\label{fig:simpleexample}
\end{figure}
To see this, we prescribe a smooth function ($\psi(x_0,y_0)$) of field lines 
 on the lower boundary of each of the
braided fields $E^{n}$, $n=1,2,  \ldots $.  We then map this function
($\psi$) along field lines of $E^{n}$ to 
 the upper boundary, i.e. find  $\psi({\bf F}^{n}(x_0,y_0))$.
This is demonstrated for a specific example in Figure~\ref{fig:simpleexample} for $E^{i}$, $i=1, \ldots,3$.
As ${\bf F}$ is applied repeatedly to any function $\psi(x_0,y_0)$, then the characteristics of the map will 
result in a stretching and contracting of $\psi$ in some regions of the domain according to the nature of 
${\bf F}$ at those locations, specifically the Lyapunov exponents of the map. After repeated application of  
${\bf F}$ any two functions will appear similar;  the apparent structure of the two functions will be determined 
by the mapping ${\bf F}$.

\begin{figure}
\begin{center}
\includegraphics[width=0.49\textwidth]{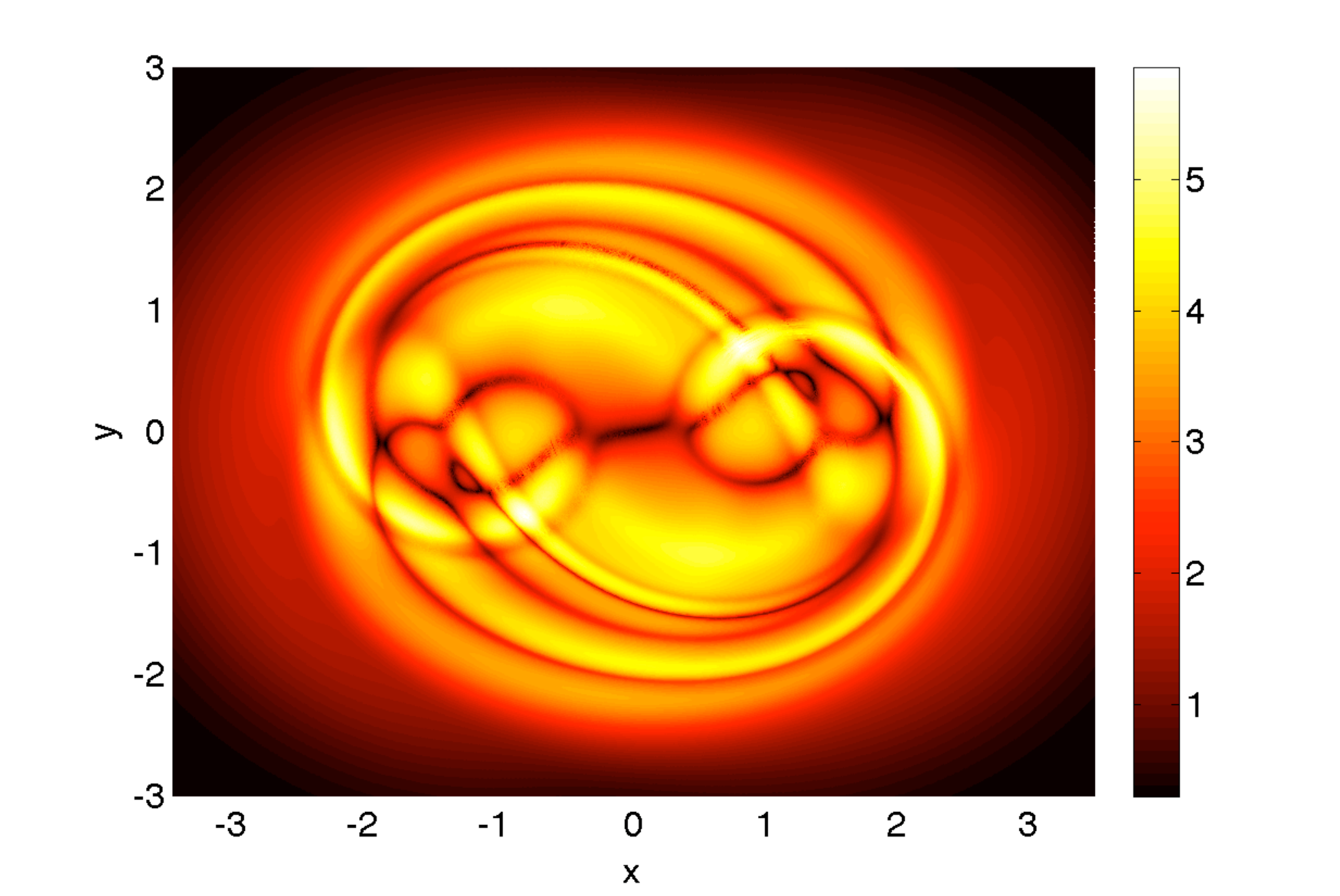}
\includegraphics[width=0.49\textwidth]{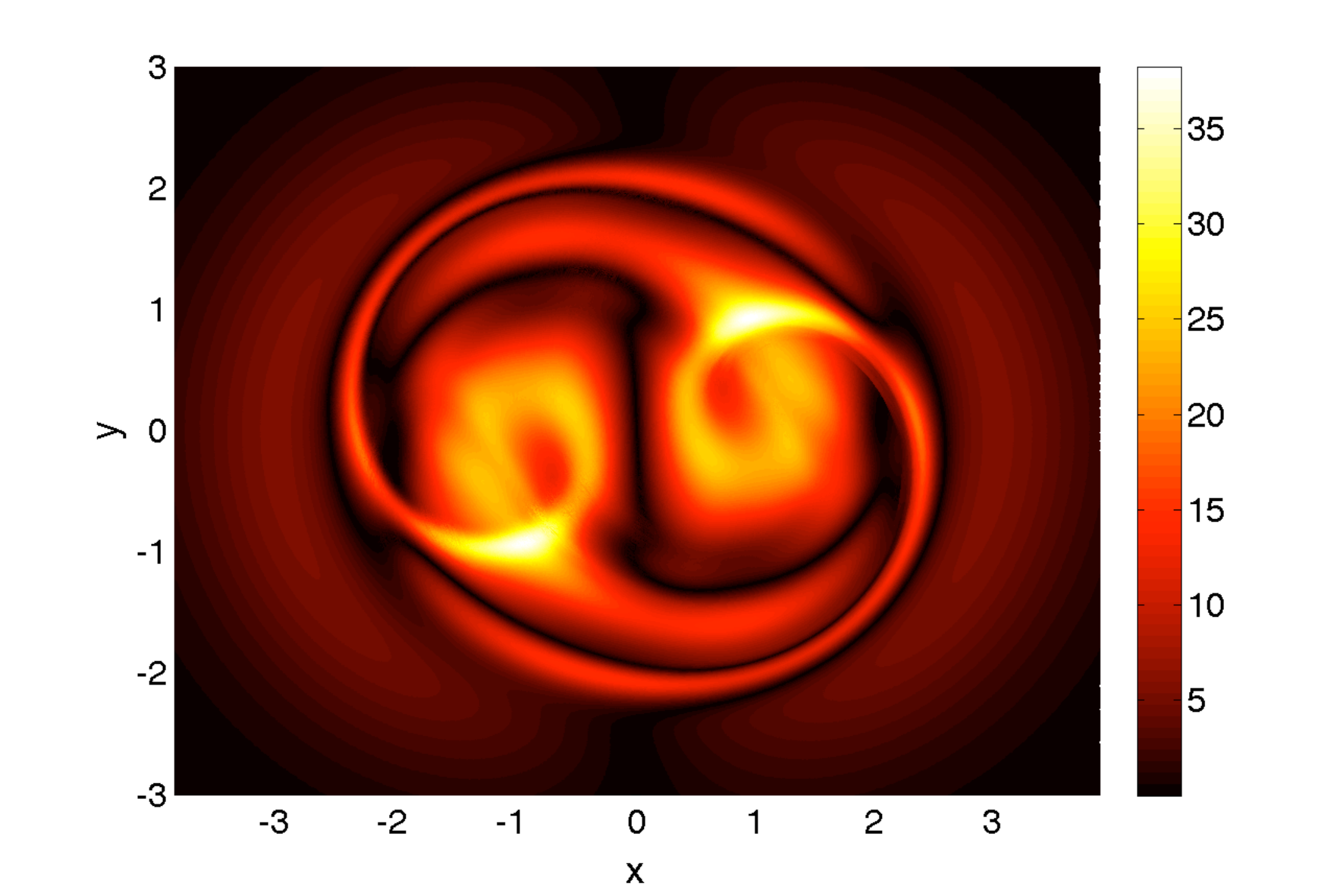}
\end{center}
\caption{({\it left}) $\textrm{log}_{10} Q$ as a function of field line position in the central
plane ($z=0$) for the field $E^{3}$ and ({\it right})
integrated parallel current along field lines,
$\vert \mathcal{J}_{\parallel}\vert $, for the same field, also in the central plane.} 
\label{fig:centre}
\end{figure}

 From this observation we deduce that the structure of both $Q$ and  $\mathcal{J}_{\parallel}$ will 
depend crucially on the plane in which they are viewed and, by the symmetry of the braid,
have the same scale on both the lower and upper boundaries.
Accordingly, $\mathcal{J}_{\parallel}$ will, for sufficiently high $n$ (about $3$ in this case),
have a filamentary distribution on the lower boundary that is reflective of the
field line mapping rather than the `underlying' nature of the parallel currents.
These properties are confirmed by illustrating $Q$ and  $\mathcal{J}_{\parallel}$ 
for $E^{3}$  in a cross-section in the middle of the domain ($z=0$), 
 thereby subtracting as much as possible the effect of the field line mapping. 
 This is shown in Figure~\ref{fig:centre} where the quantities are seen to attain 
 larger scales than on the domain boundaries and less similarity between
 $Q$ and $\mathcal{J}_{\parallel}$ is apparent.
 
 \section{Conclusions}

In this paper we have examined the link between the squashing factor $Q$ and
current structures in a class of braided magnetic fields.
The squashing factor (Titov {\it et al.} 2002) is a geometric measure for continuous 
fields that serves to identify regions where the mapping of magnetic field lines is
highly distorted.  It is used to identify quasi-separatrix layers (QSLs, Priest \& D{\'e}moulin 1995) --
features of continuous fields which provide an analogue to separatrix surfaces in fields containing
null points.  Previously, regions of high $Q$ have been considered to be a reliable indicator for the
 formation of current sheets under certain deformations on the boundaries.

Here we are interested in both the relationship between $Q$ and the current ${\bf j}$, as well
as that between $Q$ and the parallel current integrated along field lines, $\mathcal{J}_{\parallel}$.
The reason for this is that in 3D a key quantity for magnetic reconnection is the parallel 
electric field integrated along field lines.  In resistive MHD with uniform resistivity the parallel 
electric field and parallel current are related via
\[ \int E_{\parallel} \ dl = \eta \int j_{\parallel} \ dl = \eta \mathcal{J}_{\parallel}. \]

We have considered a particular class of braided magnetic fields with no net twist,
the fields being labelled $E^{n}$ ($n=1,2, \ldots$), where an increase in 
$n$  results in a field of increased complexity. 
In an earlier paper (Wilmot-Smith {\it et al.}~2009, Paper I) the field $E^{3}$ was taken as an 
initial condition in a magnetic relaxation simulation.  The field on the boundaries was
held fixed and a magnetofrictional code used to carry out an ideal relaxation toward a force-free equilibrium. 
In that process the spatial scales associated with the current, ${\bf j}$, were found to remain 
large, suggesting a smooth equilibrium corresponding to each $E^{n}$ can be found. 

In Paper I the structure of $\mathcal{J}_{\parallel}$ for $E^{n}$ was examined. This quantity 
was shown to display small spatial scales and to have an increasingly filamentary
structure with increasing $n$.
This structure was preserved in the ideal relaxation process.  
The consequence was shown to be that for a coronal field with any finite resisitivity, 
after a certain degree of braiding via photospheric motions the field will 
undergo a loss of equilibrium due to the high gradients in
 $\mathcal{J}_{\parallel}$, regardless of the current structure itself.

Here we have investigated the nature of the squashing factor $Q$ and
associated QSLs for the fields $E^{n}$.  A method was given (Appendix A)
to calculate $Q$ exactly, a property that becomes important in the analysis where 
extremely small scales in the quantity are found. 
It was shown that for the field with the least structure, $E$,  $Q$ is 
enhanced  in two layers of width $\sim 5 \%$ of the domain
width, i.e. there are two quasi-separatrix layers (QSLs) present. 
As the degree of braiding is increased, $Q$ develops an increasingly 
filamentary structure, with many narrow layers of enhanced $Q$ present.
 These are true  layers in the sense that  $Q$ drops to low values between them.
The typical width of the layers  decreases exponentially with $n$ -- for $E^{3}$ a
typical layer width is $0.01\%$ of the domain width.
Additionally, with increasing $n$ the maximum value of $Q$ exponentially increases.
 The result of the braiding process is shown to be that the domain  becomes filled with a 
 multitude of QSLs.

Despite these findings, we find that under ideal relaxation the braided fields $E^{n}$ 
do not develop any small-scale structures in the current ${\bf j}$.  
This demonstrates an alternative to the view that, in a configuration with QSLs
(outlined by regions of high $Q$) plasma motion is highly likely to lead to the 
formation of current sheets along the QSLs --
another possibility is that the magnetic field may adjust to a neighbouring 
smooth equilibrium.  This equilibrium may itself still contain QSLs.

We also considered the relationship between $Q$ and the integrated 
parallel current, $\mathcal{J}_{\parallel}$.  
In the simplest braid, $E$, the two quantities are unrelated.  With
increasing braid complexity ($n$), $Q$ and $\mathcal{J}_{\parallel}$ have some similar
characteristics;  for moderate and large $n$ both quantities are filamentary with layer-like 
regions in which high values are obtained.
However, no simple relationship has been found between the locations of maxima in 
the two quantities.

%A simple example was given (Section~5) of a 1D force-free field containing
%layer-like regions in both the squashing factor $Q$ and the current.  There the quantities were 
%shown to be unrelated.

 \newpage
 \section*{Appendix A} 
 
 Here we give a method for calculating the squashing factor in the braided fields $E^{n}$.
 For this we must obtain the mapping of field lines from the lower boundary 
of the domain to the upper boundary.  We make use of a property of the fields $E^{n}$, specifically
that $E^{n}$ is a concatenation of $n$ times the  `basic' field $E$.  

 Consider first the field $\mathbf{B}_{c} + b_{0} \hat{\bf z}$ (see Equation~\ref{eq:unittwist}), 
that is, a single toroidal flux ring imposed on the background uniform field.
The equations $\textrm{\bf{X}}\left(\mathbf{x}_{0},s\right)$ of the field line passing
through the point $\mathbf{x}_{0}=(x_{0},y_{0},z_{0})$  are obtained by integrating
\begin{displaymath}
\frac{ \partial \textrm{\bf{X}}\left(s\right)}{\partial s} = \mathbf{B} \left(
\textrm{\bf{X}}\left(s\right)\right)
\end{displaymath}
where the parameter $s$ is related to the distance $\lambda$ along field lines
by $\textrm{d}\lambda = \vert B \vert \ \textrm{d}s$.  The components of 
$\textrm{\bf{X}}\left(\mathbf{x}_{0},s\right)$ are given by
\begin{eqnarray}
\label{eq:onemap0}
\textrm{X}
 &=&  \left(x_{0} - x_{c} \right) \textrm{cos} \ \zeta  - \left(y_{0} - y_{c} \right) \textrm{sin} \ \zeta + x_{c},      \notag   \\
\textrm{Y}
 &=& \left(y_{0} - y_{c} \right) \textrm{cos} \ \zeta  + \left(x_{0} - x_{c} \right) \textrm{sin} \ \zeta + y_{c},  \\
\textrm{Z} &=& b_{0}s + z_{0},  \notag
\end{eqnarray}
where 
\begin{displaymath}
\zeta =  k \sqrt{\pi}  \frac{l}{a} 
 \textrm{exp} \left( \frac{ -\left(x_{0}-x_{c}\right)^{2}-\left(y_{0}-y_{c}\right)^{2}} { a^{2}} \right) 
  \left(
  \textrm{erf} \left(  \frac{b_{0}s+z_{0}-z_{c}}{l}   \right)  -  \textrm{erf} \left(  \frac{z_{0}-z_{c}}{l}  \right)
 \right).
\end{displaymath}

For the field $E$,  the two toroidal flux rings have been placed 
sufficiently far apart compared with their characteristic length scales that the
field in a plane equidistant b between them can be approximated as vertical, $b_{0} \hat{\bf z}$.
Accordingly, we may replace the error functions in Equations~\eqref{eq:onemap0} by $\pm 1$
and obtain an expression for the mapping of field lines from the point $(x,y)$ 
in a 2D plane (perpendicular to the field) below the single region of twist in 
$\mathbf{B}_{c} + b_{0} \hat{\bf z}$ to the point  $\mathbf{f}\left(x,y\right)$ in similar
a plane above it:
\begin{eqnarray}
%\label{eq:2dmap}
&&{\bf f}_p:   \mathbb{R}^{2} \rightarrow \mathbb{R}^{2}  \notag \\
&&\mathbf{f}_p:  (x,y) \rightarrow 
\left(\left(x - x_{c} \right) \textrm{cos} \ \xi  - \left(y - y_{c} \right) \textrm{sin} \ \xi + x_{c} , 
\left(y - y_{c} \right) \textrm{cos} \ \xi  + \left(x - x_{c} \right) \textrm{sin} \ \xi + y_{c}  \right), \notag
\end{eqnarray}
where $ p = \lbrace x_{c},y_{c},k,a,l \rbrace$ and
\begin{equation}
\label{eq:xi}
\xi = 2 \sqrt{\pi} k \frac{l}{a} \textrm{exp}
\left( \frac{ \scriptstyle-\left(x-x_{c}\right)^{2}-\left(y-y_{c}\right)^{2}} { \scriptstyle a^{2}} \right).
\end{equation}

The mapping of field lines from the lower to upper boundary of the field $E$ may then be
found by composition of ${\bf f}_{p}$ with itself using with the relevant parameters
 for each of the mappings:
\begin{equation}
\label{eq:F}
 {\bf F}(x,y) =  {\bf f}_{p_{2}} \circ {\bf f}_{p_{1}}, \notag
 \end{equation}
 where $p_{1}=\lbrace 1,0,1,\sqrt{2},2 \rbrace$ and $p_{2} = \lbrace 1,0,-1,\sqrt{2},2 \rbrace$. 
 A similar mapping, ${\bf F}^{n}(x,y)$, of the field lines from the lower to 
the upper boundary of $E^{n}$ is obtained by repeated composition of 
${\bf F}$.   For example, the field line mapping from the lower to upper 
boundary of $E^{3}$  is  ${\bf F}^{3}\left(x,y\right) = 
{\bf F} \circ {\bf F}^{2} = {\bf F} \circ {\bf F} \circ {\bf F}$.

The squashing factor $Q$ for $E^{n}$ can be found directly from
the Jacobian ($\mathbf{DF}^{n}$) of the mapping $\mathbf{F}^{n}$:  
\[ Q = \frac{\vert \vert {\bf DF}^{n} \vert \vert ^{2}}{\textrm{det}\left({\bf DF}^{n}\right)}\]
where {\small $\vert \vert  \cdot \vert \vert$} denotes the entrywise $p=2$~norm. 
In other words, letting
\[{\bf DF}^{n} = \left(
\begin{array}{ll}
\frac{\partial X}{\partial x} &
\frac{\partial X}{\partial y} \\
\frac{\partial Y}{\partial x} &
\frac{\partial Y}{\partial y} \\
\end{array}
\right)
= \left( \begin{array}{ll}
a & b \\
c & d
\end{array} \right)
\]
say, then
\[Q = \frac{a^{2}+b^{2}+c^{2}+d^{2}}{ad-bc}.\]
For the fields $E^{n}$ the determinant of the Jacobian mapping ${\bf DF}^{n}$
is exactly unity ($\textrm{det}({\bf DF}^{n})=ad-bc=1$) since on both boundaries the field
is the constant field $b_{0} {\hat{\bf{z}}}$.
 This method determines $Q$ in the initial states for $E^{n}$.  Since in the
 ideal relaxation towards a force-free equilibrium the boundaries are held fixed, the 
 structure of  $Q$ on the boundaries of the domain is preserved throughout the experiments.

  Using the same approximation as above, we 
are able to find a closed form expression for the integrated parallel current along field lines
 in the initial states for $E^{n}$.  
Taking only a single region of  toroidal twist, the field line with initial location $(x,y)$ in a 
plane below the region has a total integrated parallel current of
$$\mathcal{J}_{\parallel, \begin{smallmatrix}
\textrm{one} \\
\textrm{twist}
\end{smallmatrix}} = 
\int_{\begin{smallmatrix}
\textrm{one} \\
\textrm{twist}
\end{smallmatrix}} 
j_{\parallel} dl
= \frac{2b_{0}}{\mu  a} \left(a^{2} - \left(x-x_{c}\right)^{2}
- \left(y - y_{c}\right)^{2}\right) \xi,$$
where $\xi$ is defined in Eqn.~\eqref{eq:xi}.
The total integrated parallel current along field lines for $E^{n}$ may 
be found using repeated composition of this expression in a manner 
similar to that used for $Q$. Note that in Paper I $\mathcal{J}_{\parallel}$
was evaluated directly via numerical integration.

\newpage

%\vspace*{1cm}
\noindent
{\bf Bibliography} \\

\vspace*{0.15cm} \noindent
{Aulanier}, G., {Pariat}, E., {D{\'e}moulin}, P., {\it Current sheet formation in quasi-separatrix layers 
and hyperbolic flux tubes}, A\&A, {\bf 444}, 961-976 (2005).  

\vspace*{0.15cm} \noindent
Bineau, M., {\it On the Existence of Force-Free Magnetic 
Fields}, Comm. Pure Appl. Math., 27, 77 (1972)

\vspace*{0.15cm} \noindent
{Craig}, I.~J.~D. and {Sneyd}, A.~D., 
{\it {The Parker Problem and the Theory of Coronal Heating}}
Solar Physics, {\bf 232}, 4-621(2005). 

\vspace*{0.15cm} \noindent
{Fletcher}, L.,  {L{\'o}pez Fuentes}, M.~C., {Mandrini}, C.~H.,  {Schmieder}, B., {D{\'e}moulin}, 
P., {Mason}, H.~E., {Young}, P.~R.,  {Nitta}, N., {\it A Relationship Between Transition Region 
Brightenings, Abundances, and Magnetic Topology} Solar Physics, {\bf 203} 255-287 (2001). 

\vspace*{0.15cm} \noindent
{Galsgaard}, K. and {Nordlund},  {\AA},
{\it Heating and activity of the solar corona 1. Boundary shearing of an initially 
homogeneous magnetic field}, JGR, {\bf 101}, 13445-13460 (1996) 

\vspace*{0.15cm} \noindent
Galsgaard, K., Titov, V.S., Neukirch, T., {\it Magnetic pinching of hyperbolic
flux tubes. II. Dynamic numerical model} ApJ {\bf 595} 506-516 (2003). 
 
 \vspace*{0.15cm} \noindent
{Longcope}, D.~W.,  {Strauss}, H.~R., {\it The form of ideal current layers in line-tied magnetic fields}, 
ApJ,  {\bf 437} 851-859 (1994). 

\vspace*{0.15cm} \noindent
Longbottom, A.W, Rickard, G.J., Craig, I.J.D., and Sneyd, A.D., {\it Magnetic flux braiding:Force-free equilibria and current sheets}, 
 ApJ {\bf 500} 471-482 (1998). 

\vspace*{0.15cm} \noindent
Milano, L.~J., Dmitruk, P., Mandrini, C.~H., Gomez, D.~O., D{\'e}moulin, P.,
{\it Quasi-Separatrix Layers in a Reduced Magnetohydrodynamic Model of a Coronal Loop} 
ApJ, {\bf 521} 889-897 (1999). 
  
\vspace*{0.15cm} \noindent  
{Parker}, E.~N., {\it Topological Dissipation and the Small-Scale Fields in Turbulent Gases},
ApJ, {\bf 174} 499 (1972.) 

\vspace*{0.15cm} \noindent
{Pontin}, D.I., Hornig, G., Wilmot-Smith, A.L., Craig, I.J.D.,
{\it Lagrangian relaxation schemes for calculating force-free magnetic fields, and their limitations},
ApJ,  {\bf 700}, 1449 (2009)

\vspace*{0.15cm} \noindent
{Priest}, E.~R.,  {D{\'e}moulin}, P., {\it Three-dimensional magnetic reconnection without null points. 
1. Basic theory of magnetic flipping}, JGR, {\bf 100} 23443-23464 (1995). 

\vspace*{0.15cm} \noindent
{Schmieder}, B., {Aulanier}, G., {Demoulin}, P., {van Driel-Gesztelyi}, L., {Roudier}, T., {Nitta}, N., 
{Cauzzi}, G., {\it Magnetic reconnection driven by emergence of sheared magnetic field},
A \& A {\bf 325} 1213-1225 (1997). 

\vspace*{0.15cm} \noindent
{Titov}, V.~S., {\it Generalized Squashing Factors for Covariant Description of Magnetic Connectivity 
in the Solar Corona}, ApJ, {\bf 660}, 863 (2007). 

\vspace*{0.15cm} \noindent
{Titov}, V.~S. and {Galsgaard}, K. and {Neukirch}, T.,
{\it  Magnetic Pinching of Hyperbolic Flux Tubes. I. Basic Estimations}, ApJ,
{\bf 582}, 1172 (2003).

\vspace*{0.15cm} \noindent
{Titov}, V.~S., {Hornig}, G.,  {D{\'e}moulin}, P.,
    {\it Theory of magnetic connectivity in the solar corona}, JGR (Space Physics)
{\bf 107}, 1164 (2002). 

\vspace*{0.15cm} \noindent
{Titov}, V.~S., {Forbes}, T.~G., {Priest}, E.~R., {Mikic}, Z., {Linker}, J.~A.,
  {\it Slip-Squashing Factors as a Measure of Three-Dimensional Magnetic Reconnection},
  ApJ, {\bf 693}, 1029 (2009). 
  
  \vspace*{0.15cm} \noindent
   {Van Ballegooijen}, A.~A., 
{\it Electric currents in the solar corona and the existence of magnetostatic equilibrium},
  ApJ, {\bf 298}, 421-430 (1985) 
  
  \vspace*{0.15cm} \noindent
  {Wilmot-Smith}, A.~L.,  {Hornig}, G.,  {Pontin}, D.~I.,
  {\it Magnetic Braiding and Parallel Electric Fields},
 ApJ, {\bf 696}, 1139 (2009).

\end{document}